\documentclass[conference]{IEEEtran}
\usepackage{tikz}
\usepackage{amsmath}
\usepackage{url}
\usepackage{doi}
\usepackage{cite}
\usepackage{xspace}
\usepackage{mathptmx}
\usepackage{hyperref}
\usepackage{cleveref}
\usepackage{balance}

\def\BibTeX{{\rm B\kern-.05em{\sc i\kern-.025em b}\kern-.08em
    T\kern-.1667em\lower.7ex\hbox{E}\kern-.125emX}}

\newcommand{\etal}{\textit{et al.}\xspace}

\title{Continuous Autonomous Refactoring:

A Research Roadmap for AI-Driven Code Quality Maintenance}

\author{
\IEEEauthorblockN{Xin Sun, Daniel Ståhl, Kristian Sandahl, Christoph Kessler}
\IEEEauthorblockA{
Department of Computer and Information Science\\ Linköping University, Sweden\\
\{xin.sun, daniel.stahl, kristian.sandahl, christoph.kessler\}@liu.se }}

\begin{document}
\maketitle

\begin{abstract}
    Large language models have shown promising capabilities in code refactoring, but existing approaches remain limited to method-level tasks. In this paper, we envision LLM-based refactoring as a continuous component of software maintenance rather than a tool invoked only for occasional manual refactoring. Under this vision, AI agents continuously monitor, evaluate, and improve codebases against explicit and evolving notions of software quality. We present a roadmap organized around five dimensions: the multi-objective optimization problem, quality definition and evaluation, multi-timescale integration of heterogeneous signals, architecture and design pattern, and trust in autonomous refactoring. We further identify integration into continuous delivery pipelines and cost considerations as cross-cutting concerns. For each dimension, we analyze the underlying challenges and pose open research questions. These dimensions define a research agenda for advancing autonomous refactoring from isolated code improvements to system-level quality maintenance.
\end{abstract}

\section{Introduction}
\label{intro}

The emergence of large language models (LLMs) has brought substantial changes to software engineering. These models demonstrate strong capabilities in generating and modifying source code from natural language prompts \cite{wangAreview2023}. Such capabilities lead to a growing interest in using LLMs not only to implement new features but also to improve existing codebases through code refactoring, aiming to improve code quality and manage technical debt \cite{OpenjaMAKA22}.

Recent work suggests that LLMs can effectively perform code refactoring on relatively straightforward tasks, reducing local complexity, repairing specific defects, and improving performance efficiency while preserving functional behavior \cite{martinezSoftwareRefactoringResearch2026}. However, in more complex settings, LLM-based refactoring often encounters failures. At the repository level, generated changes may unintentionally violate functional correctness or cause substantial degradation in certain quality dimensions while improving others \cite{sun2025qualityassurancellmgeneratedcode}.

Consider a team maintaining a large-scale enterprise system spanning several million lines of code across hundreds of modules. Static analyzers flag thousands of code smells, but the team knows many are intentional: a deeply nested function preserves a real-time latency constraint, and a duplicated validation block remains in two services because extracting a shared library previously caused integration failures. An LLM-based tool might simplify a single method in isolation, but no existing system can determine which warnings should be addressed, balance a cross-module refactoring against an upcoming release deadline, or recognize that a redundant adapter layer preserves backward compatibility with an external API. This gap highlights the need to reassess how close we are to system-level automated software refactoring and what research methods are required to achieve it.

In this paper, rather than asking how to ensure quality in AI-generated code, we pose a different question: \emph{How could AI agents continuously and autonomously maintain code quality against an explicit, evolving specification of ``good''?} This reframing shifts quality from a constraint on code generation to a primary and ongoing objective. In this view, continuous autonomous refactoring becomes the default mode of software maintenance. AI agents monitor, evaluate, and improve codebases against evolving quality criteria, instead of a one-time intervention applied to isolated code snippets. This paradigm shift raises research questions that go beyond current capabilities and form the basis of the roadmap we present in this paper.

We present a research roadmap toward realizing the vision of continuous autonomous refactoring, organized around five primary dimensions: \emph{the multi-objective optimization problem}, \emph{the problem of definition and evaluation}, \emph{multi-timescale integration of heterogeneous signals}, \emph{architecture and design pattern challenges}, and \emph{the trust problem}. Beyond these core dimensions, we further discuss \emph{integration into continuous delivery pipelines} as a cross-cutting concern and \emph{cost considerations for industry adoption} as a practical concern for industry adoption. Each dimension addresses a distinct group of challenges that must be resolved before autonomous refactoring can scale beyond isolated tasks. Together, they define the research questions that can be taken by the community to make this vision a reality.

\section{Background and Related Work}
\label{back}
\subsection{Code Refactoring}
Code refactoring is an essential practice in software engineering. Unlike automated program repair, which aims to modify programs to fix bugs, code refactoring improves the internal structure without changing its behavior \cite{fowler2018refactoring}. Traditional refactoring relies heavily on developer experience and manual operations, requiring a profound understanding of software architecture and code quality \cite{BaqaisA20}. However, as software systems grow in complexity, manual refactoring becomes a time-consuming and error-prone task, leading many quality improvements to be deferred \cite{karabiyikRefactorGPTChatGPTbasedMultiagent2025}. The accumulation of deferred quality improvements leads to technical debt \cite{kruchten2012TD}, which has been extensively studied as a critical challenge in software evolution. Importantly, technical debt is broader than just postponing to deal with code smells. It also includes invisible issues such as architectural debt and technological gaps, many of which cannot be fully revealed by static analyzers. Managing technical debt systematically requires not only identifying problematic code but also prioritizing and planning remediation efforts within broader project constraints \cite{LiAL15}. Existing machine learning approaches have been effective in technical debt detection and classification, but they often struggle to capture the nuanced semantics and context embedded in technical debt descriptions \cite{daniel2020detecting}. LLMs provide new opportunities for automating parts of this process and have shown improved performance on related tasks \cite{shivashankarBEAConTDClassifyingTechnical2025}.

\subsection{Related Work}
The vision of autonomous software maintenance is not entirely new. Kephart and Chess \cite{KephartChess03} articulated the vision of autonomic computing, in which systems manage themselves through self-configuration, self-optimization, self-healing, and self-protection. Cheng \etal\cite{ChengDLIMGGKLMSWY09} subsequently established a research roadmap for software engineering of self-adaptive systems, identifying key challenges in modeling, requirements, and verification.  The current pursuit of LLM-driven autonomous refactoring can be understood as a modern instantiation of these earlier visions, now enabled by the language understanding and generation capabilities of LLMs. Some studies argue that the IDE is the most practical environment for AI-assisted refactoring, combining static analysis tools with human-centered mechanisms to enable trustworthy adoption \cite{Bellur2025IDE}. However, achieving truly autonomous refactoring requires reasoning beyond the IDE workflow, incorporating repository and system-level context, global dependencies, and long-term evolution. 

Unlike \cite{Ivers2020Nextgeneration}, which focuses on feature-level isolation, our roadmap considers refactoring at a much larger scale and emphasizes leveraging heterogeneous data sources to capture system constraints and design intent. Kr\"{u}ger \etal\cite{kruger2023vision}  argued that current tools lack the ability to capture change intent and design motivation, calling for future software engineering tools to explicitly model the latent change intentions in software evolution.

Recent studies have explored how developers interact with LLMs during refactoring tasks and how effectively LLMs can identify refactoring opportunities and generate solutions \cite{chavanAnalyzingDeveloperChatGPTConversations2024, LiuJZNLL25, shirafujiRefactoringProgramsUsing2023a}. While these studies suggest promising potential, they also show that the generated refactorings may introduce new defects, hallucinations, or syntax errors, especially in repository-level contexts, requiring additional human verification. To mitigate these issues, researchers have proposed approaches that integrate LLMs with traditional development tools, such as static analysis and IDE environments, to support more reliable method-level refactoring \cite{pomian2024nextgeneration}. Other frameworks adopt agent-based workflows and multi-agent architecture that coordinate planning, code analysis, patch generation, and validation in an end-to-end process \cite{zhangRFR24, WuM0G0ZZ024}. Empirical studies suggest that such systems can improve modularity and automation, but maintaining functional correctness and handling large-scale project contexts remain challenging.

Overall, existing studies highlight the potential of LLM-based refactoring while exposing challenges regarding refactoring quality and long-context understanding at the system level. Some reports show high rates of hallucinated or syntactically invalid suggestions \cite{pomian2024nextgeneration}. Additionally, model performance often degrades when addressing complex problems or large-scale codebases. At the same time, LLM capabilities are evolving rapidly, and all empirical results discussed in this section must be understood in the context of the specific model generation studied. Separating generalizable characteristics of LLM behavior from transient capability gaps remains an open methodological challenge. Benchmarks such as SWE-bench \cite{jimenez2024swebench}, which evaluate LLM agents on real-world GitHub issues, illustrate both the rapid progress and the remaining gaps: resolution rates have improved dramatically across model generations, yet consistent system-level reasoning remains elusive. Therefore, in this paper, we aim to look past limitations that may vanish with stronger models and to focus on structural challenges that are likely to persist regardless of incremental capability gains.

\section{Roadmap}
\label{roadmap}

The dimensions presented in this roadmap were derived through a thematic analysis\cite{Braun2006} of the limitations and open challenges reported across the studies reviewed in \autoref{back} and the work of \cite{sun2025qualityassurancellmgeneratedcode}. We first examined the research questions, conclusions, and discussion sections of the selected studies. Statements describing reported challenges and future research directions were extracted and assigned initial codes representing underlying concerns about autonomous refactoring. Then, we iteratively compared and grouped the codes into semantically related themes. This process yielded five dimensions that capture recurring perspectives on future challenges in autonomous refactoring research and two practical concerns for industry adoption as presented in the following subsections. 

\begin{figure*}[ht]
    \centering
    \includegraphics[width=0.65\linewidth]{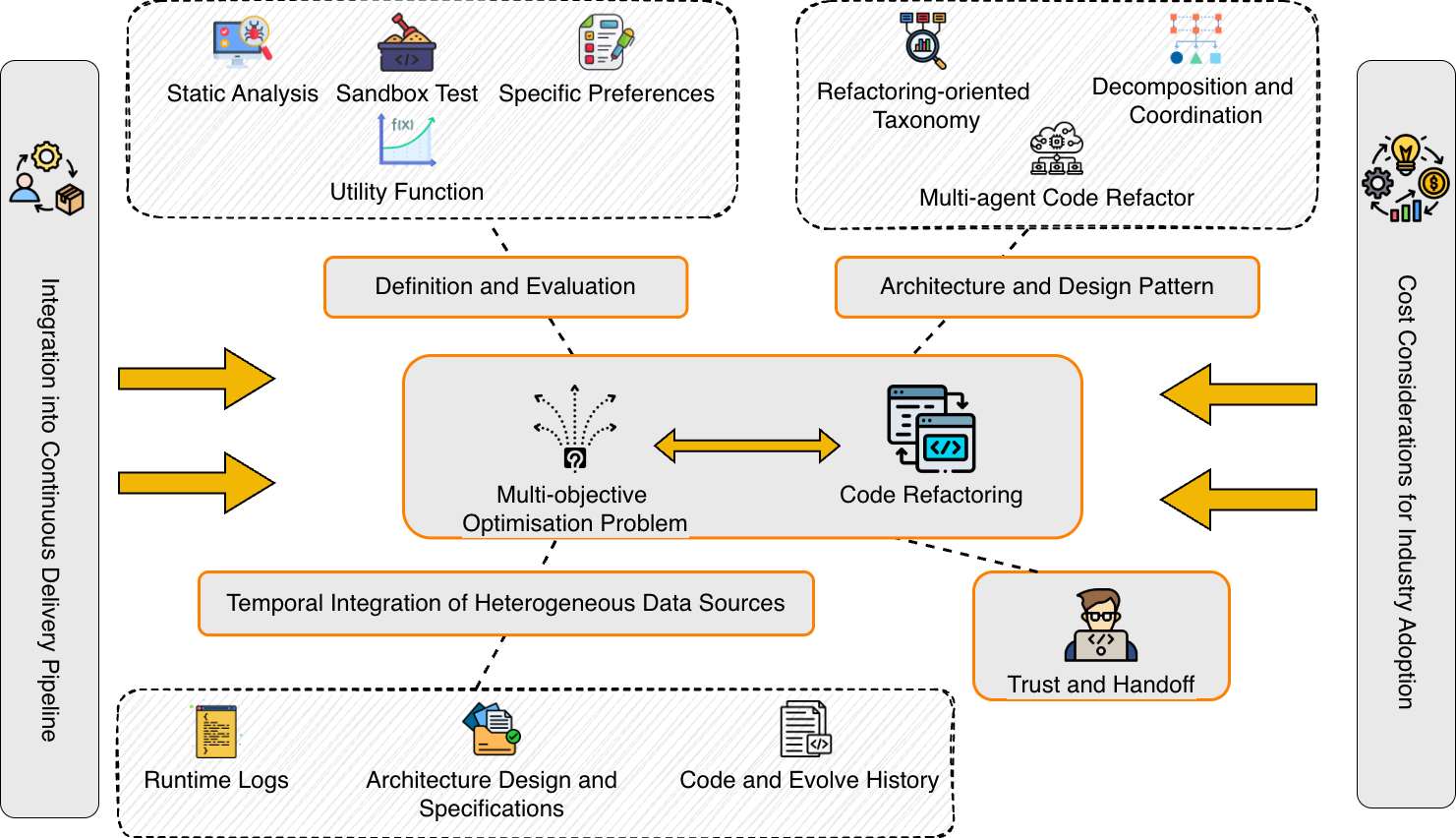}
    \caption{Research roadmap for autonomous code refactoring.}
    \label{fig:structure}
\end{figure*}

\autoref{fig:structure} shows the overview of the roadmap. For each dimension, we present evidence of the problem, articulate why it constitutes a research challenge rather than a purely engineering problem, identify the constraints and desirable properties of potential solutions, and pose concrete open research questions. These dimensions are not intended to be exhaustive; rather, they represent the most prominent and structurally distinct challenges that emerge when moving from isolated method-level refactoring to continuous system-level autonomous refactoring.

\begin{figure}
    \centering
    \includegraphics[width=\linewidth]{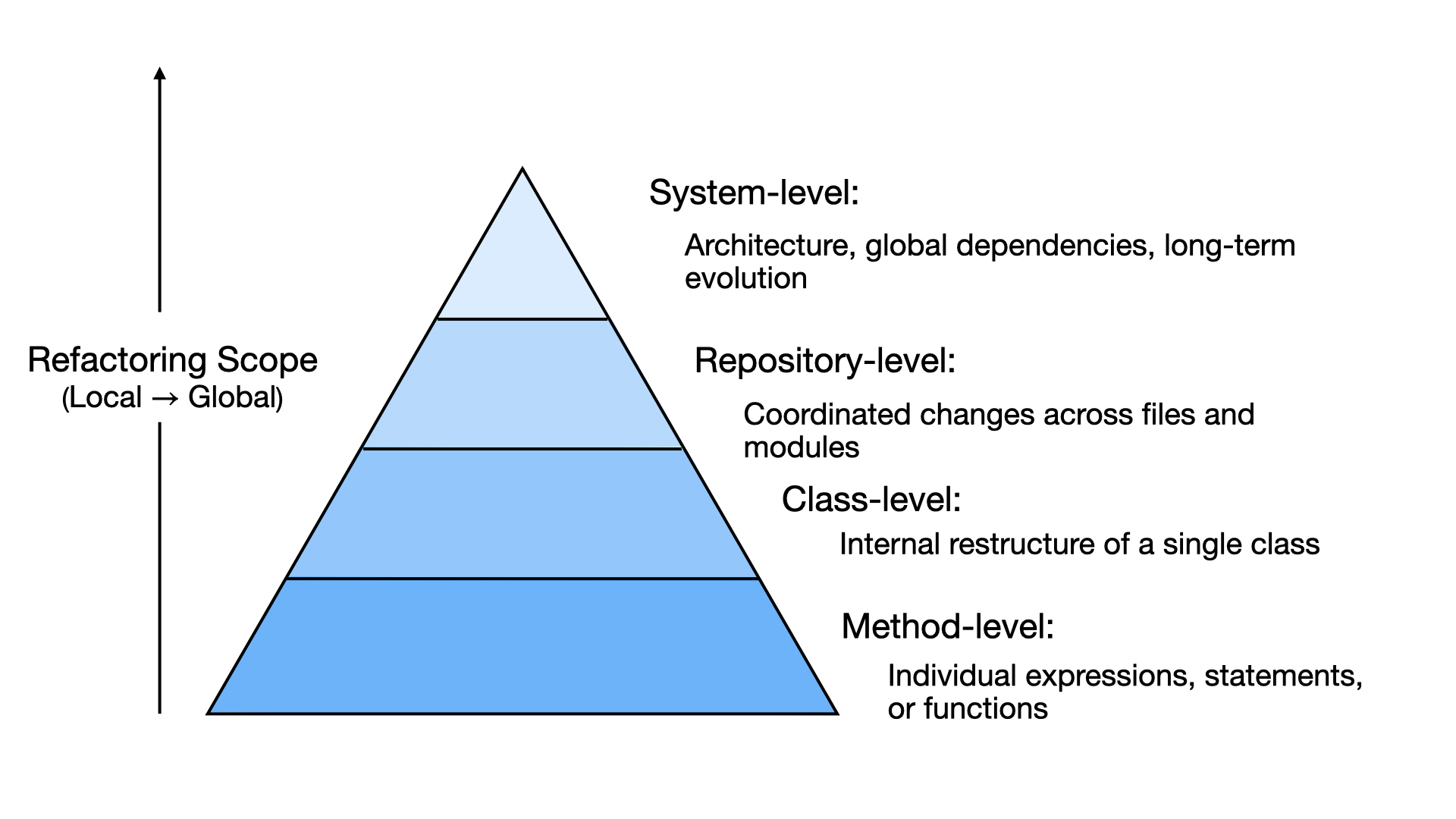}
    \caption{Refactoring scope hierarchy.}
    \label{refactoring}
\end{figure}

Additionally, to reason clearly about these challenges, we distinguish four levels of refactoring scope, inspired by \cite{martinezSoftwareRefactoringResearch2026} as shown in \autoref{refactoring}.

To enable autonomous code refactoring, an underlying concern across several dimensions is \emph{intent communication}. Much of what human engineers consider ``good code'' rests on implicit knowledge about conventions, trade-offs, and domain constraints that are rarely documented. When an AI agent violates these implicit preferences, the result is perceived as poor quality, even if the specification provided to the agent may never have specified those preferences. Addressing this gap requires new mechanisms that allow software engineers to communicate their intent to AI agents in actionable terms, so that agents can grasp contextual constraints and weigh trade-offs accordingly. Developing such mechanisms remains an open research question, to which this paper contributes a roadmap and an initial perspective.

\subsection{The Multi-objective Optimization Problem}
\label{mop}
Modern software is a complex system of interdependent, evolving components. Single-objective refactoring at the system scale often harms other quality dimensions, a phenomenon particularly evident in cross-module refactoring \cite{sun2025qualityassurancellmgeneratedcode}. From a computational perspective, automated refactoring is a constrained multi-objective optimization problem (MOP) \cite{ramirezSurveyManyobjectiveOptimisation2019}. Given the inherent conflicts between objectives, improving one often degrades another, meaning no single solution optimizes all software quality metrics simultaneously. While local optima are easily found in code snippets, the complex search space created by interacting quality metrics at the system scale requires global constraint awareness to avoid regressive evolution. 

However, framing refactoring as an MOP does not mean that existing optimization techniques can directly solve the problem. Software systems contain latent constraints that are not fully known upfront and must be revealed and refined over successive refactoring cycles. At the same time, many commonly used quality indicators are imperfect proxies for global properties, and improvements on these metrics can fail to translate into better system-level quality. These uncertainties shift the core challenge away from applying existing multi-objective optimization techniques and toward defining objective functions and constraints that reflect real-world software quality at the system level. The resulting search space is large and discontinuous, and the interactions among quality dimensions remain poorly understood at scale. 

One illustrative direction is to adopt the decomposition and coordination ideas of MOEA/D, which is a multiobjective evolutionary algorithm based on decomposition \cite{zhang2007moea}, to autonomous refactoring. A global agent sets refactoring boundaries and Pareto-front priorities based on developer preferences and established quality standards, while system-level goals are decomposed into sub-tasks executed by specialized agents. Real-time feedback would connect local modifications back to global objectives, allowing the system to detect trade-offs and resolve conflicts across quality dimensions. The open research questions, however, go deeper: \textit{How can the trade-off between quality dimensions be characterized at the repository and system levels? What feedback mechanisms are needed to detect when local improvements degrade global properties? How should the search for possible refactoring actions be organized so that system-level optimization remains manageable?}

\subsection{The Problem of Definition and Evaluation}
\label{daf}
Existing research on refactoring largely focuses on functional correctness and typically evaluates non-functional quality using static analysis tools and manual reviews \cite{wangAreview2023}. However, optimizing specific static metrics does not necessarily improve overall code quality. Evidence suggests that LLMs may increase memory usage to reduce runtime or over-partition code to lower local complexity, thereby increasing system-level coupling \cite{sun2025qualityassurancellmgeneratedcode}. Furthermore, developers and organizations often hold subjective preferences regarding what constitutes ``good code'' \cite{surucuEstablishingKeyPerformance2020}. The absence of evaluation frameworks that reconcile structured metrics with contextual preferences makes automated systems prone to local optima or the creation of unmaintainable systems. 

Simply adding more metrics or running more tests cannot solve this problem. Many quality attributes are emergent and cannot be fully captured by static analyzers, and trade-offs between attributes are common and often unavoidable. Thus, evaluations must support decision-making under incomplete observations and evolving, sometimes conflicting, preferences, rather than relying on a fixed set of universal rules. 

Thus, the core difficulty of this problem is not engineering but conceptual: how to formalize subjective, context-dependent notions of quality into representations that autonomous agents can act on. Any viable solution must combine structured metrics with specific preferences, adapt as projects evolve, and detect regressions across quality dimensions. One possible direction is an evaluation framework that integrates static analysis, sandbox testing, and organizational preferences using utility functions \cite{RadulescuMRN20} to support quantitative quality decisions. 

In such a framework, static metrics and sandbox logs would be translated into comparable utility scores reflecting project priorities, and developers could adjust weights according to their needs. By aggregating these weighted utilities, the system could identify refactoring options where expected benefits outweigh associated costs. Critically, before any non-functional quality improvement is enacted, the evaluation framework must first verify that the proposed refactoring preserves functional behavior. Existing test suites provide a practical guardrail: refactoring candidates that break tests can be automatically
rejected, ensuring that quality gains do not come at the cost of behavioral regressions. Regardless of the implementation, key research questions remain: \textit{How can subjective quality preferences be elicited, formalized, and kept up to date as projects evolve? What is the right level of granularity for quality specifications? How should conflicts between different stakeholders' preferences be resolved within such an evaluation framework?}

\subsection{Multi-timescale Integration of Heterogeneous Signals}
\label{time}

Static tools provide a snapshot of code quality and capture only a limited aspect of continuous software evolution. In agile development, quality preferences are often hidden within temporal archives such as logs, test results, and commit histories \cite{tahirSystematicMappingStudy2012}. Such data explain seemingly irrational designs, such as redundant code maintained for legacy compatibility or historical bug fixes. Moreover, different signals emerge at varying frequencies, from static metrics computed on each commit to real-time production loads, to customer feedback that arrives on much longer timescales. Autonomous refactoring must synchronize these multi-speed information sources to maintain system stability after refactoring. Yet using heterogeneous data in refactoring requires more than collecting it. 

The key challenge is reasoning about the intent and constraints behind the data at multiple abstraction levels \cite{kruger2023vision}. This requires both spatial awareness of system structure and dependencies and temporal awareness of how the project evolved and why earlier decisions were made. LLMs offer promising capabilities for processing unstructured data, but extracting actionable constraints from diverse sources remains an unsolved problem. In addition, mechanisms are needed to detect pseudo-optimizations, which appear beneficial locally but degrade quality elsewhere, so that the refactoring process remains robust across successive iterations. More specifically, the open questions include: \textit{How should heterogeneous temporal data be represented and prioritized for refactoring decisions? What architectures can support real-time integration of multi-speed data streams? How can implicit design rationale be reliably extracted from historical artifacts such as commit histories, bug reports, code reviews, and operational logs?}

\subsection{Architecture and Design Pattern Challenges}
\label{arc}

Building a practical autonomous refactoring system raises challenges in both model selection and architecture design. Selecting a model based solely on aggregate benchmark performance is insufficient, as different models exhibit distinct capability profiles that must be matched to task requirements. For example, some models handle long context better, while others perform more reliably on structured reasoning tasks \cite{hou2024LLMsforSE}. However, the field currently lacks a taxonomy that categorizes models according to these specialized capabilities, making principled model selection difficult.

Software refactoring is not a single-step generation task. It involves system-level reasoning, constraint checking, and coordination across components, which may exceed the capability of a single agent. Multi-agent architectures, in which different agents handle different subtasks, offer a promising direction, but introduce new challenges in communication, coordination, and conflict resolution \cite{WuM0G0ZZ024}.

To address these challenges, the field would benefit from an evaluation framework grounded in model performance on refactoring tasks, rather than relying on existing rankings based primarily on general coding ability for model selection. If multi-agent approaches are to scale, they must include interpretable coordination and conflict-resolution mechanisms that reconcile competing edits across dependencies, semantics, and quality objectives. These considerations lead to several open questions: \textit{What capabilities should be used to classify models so that appropriate models can be selected for different refactoring sub-tasks? What coordination mechanisms can resolve conflicts when multiple agents propose competing changes? How should the system architecture itself adapt as model capabilities evolve rapidly?}

\subsection{The Trust Problem}
\label{trust}
Automation does not eliminate the need for human involvement; rather, the nature of that involvement must evolve. A reliable system must define clear boundaries of autonomy and determine when human intervention is required, based on task risk and output uncertainty \cite{taker2025humanintheloop}. Granting full autonomy in high-risk scenarios can lead to serious failures, while escalating too many low-risk cases to humans undermines efficiency \cite{abraha2025}. The deeper question is not how to route decisions to humans, but what it takes to build sufficient trust so that the scope of autonomous action can grow over time.

Building such trust requires mechanisms through which the refactoring system's behavior can be incrementally evaluated and verified. When handoff to human developers does occur, the interaction should itself be informative: agents should provide clear rationales, expected benefits, and potential side effects, enabling developers to shift from line-by-line code inspection to validating intent and decision logic. These handoff interactions should also be recorded as temporal feedback, letting the system learn implicit constraints and refine its preference model from human responses. Thus, the open questions are: \textit{How can the trustworthiness of an autonomous refactoring system be measured in practice? What evidence and explanation mechanisms are needed for developers to trust autonomous refactoring decisions?}

\subsection{Cross-Cutting and Practical Considerations}

\paragraph{Integration into Continuous Delivery Pipelines}

An important challenge that spans all of the dimensions above is integrating autonomous refactoring into continuous delivery pipelines while ensuring that its outcomes can be reliably assessed. Modern software organizations increasingly adopt continuous integration, continuous delivery, and continuous deployment practices, and require any autonomous refactoring system to fit into these workflows. While prior work emphasizes that every change should be deployable and verifiable through automation \cite{HumbleFarley10}, it remains an open question where autonomous refactoring should be placed within such pipelines, whether before commit, after integration, or as a parallel process. Additionally,  continuous software engineering practices are still evolving, which introduces extra uncertainty into this integration challenge \cite{RodriguezHLTSEKKVO17}. 

This challenge is further compounded by the need to verify not only functional correctness but also the non-functional quality improvements that refactoring is intended to achieve. Existing research in continuous practices has identified verification as a persistent bottleneck, and autonomous refactoring introduces additional uncertainty. The open research questions include: \textit{How should autonomous refactoring be positioned within continuous delivery pipelines? How can the feedback loops inherent in continuous delivery be leveraged to improve the refactoring system itself?}

\paragraph{Cost Considerations for Industry Adoption}
Finally, any roadmap toward continuous autonomous refactoring must account for cost. System-level refactoring can require repeated LLM calls for analysis, generation, and verification, and the resulting inference and infrastructure costs scale with repository size and iteration frequency \cite{WuM0G0ZZ024}. Whether continuous autonomous refactoring is practical, therefore, depends on whether the value it creates, through reducing technical debt and maintenance effort, outweighs the costs of keeping it in operation. This constraint also motivates cost-aware scheduling, reuse of intermediate results across cycles, and prioritization of actions with high expected impact. Beyond this, continuous automation may also have environmental implications due to increased energy consumption and may gradually reshape development workflows across the field.

\section{Conclusion}
\label{conclusion}
This paper argues for a paradigm shift in how software quality should be managed in the age of LLM-driven development. Rather than treating refactoring as an occasional, manually initiated intervention, we propose that continuous autonomous refactoring should be integrated into software maintenance as a persistent process, in which LLM-based agents monitor, evaluate, and improve codebases against an explicit and evolving specification of what constitutes ``good''. Under this view, quality is no longer a constraint on code generation but a primary, ongoing objective.

The roadmap outlined in this paper identifies the research challenges that need to be addressed to realize this vision. A key insight is that the strengths of LLMs in code generation can be combined with the verification capabilities of established engineering tools. This combination enables improving local code quality while preserving global architectural consistency. In addition, incorporating temporal data and utility-based evaluation helps ensure that automated refactoring remains aware of project history and aligned with specific priorities of the team. The earlier traditions of autonomic computing \cite{KephartChess03} and self-adaptive systems \cite{ChengDLIMGGKLMSWY09} laid the conceptual groundwork; the capabilities of modern LLMs now make a renewed and more ambitious pursuit of this vision both timely and feasible.

We call on the research community to take up these questions. Tools and model capabilities are advancing rapidly, but the conceptual and methodological foundations for system-level autonomous refactoring still lag behind. Progress on these dimensions would not only advance automated software maintenance but also reshape how human developers collaborate with AI systems. It points toward adaptive software that evolves continuously and manages its own quality over time. We plan to pursue this agenda in our future work, and we invite the broader community to help define the methods, tools, and evaluation frameworks needed to bring continuous autonomous refactoring from vision to practice.

\balance
\bibliographystyle{IEEEtran}
\bibliography{reference}

\end{document}